\documentstyle[11pt,iau185,twoside,epsf]{article}

\begin{document}

\title{Frequency analysis of $\delta$~Scuti and RR~Lyrae stars in the 
       OGLE-1 database}
\author{P.~Moskalik}
\affil{Copernicus Astronomical Centre, Warsaw, Poland}
\author{E.~Poretti}
\affil{Osservatorio Astronomico di~Brera, Merate, Italy}

\begin{abstract}
We discuss the results of a systematic search for multi\-periodic pulsations
in Galactic Bulge $\delta$~Scuti and RR~Lyrae stars. Six "normal" double-mode
variables pulsating in two radial modes have been identified (5
$\delta$~Scuti-type and 1 RR~Lyrae-type). In 37 RR~Lyrae stars secondary
periodicities very close to the primary pulsation frequency have been
detected.  These periodicities correspond to nonradial modes of oscillation. 
They are found in $\sim\! 23\%$ of RRab and in $\sim\! 3\%$ of RRc variables 
of our sample. 
\end{abstract}

 
\section{Introduction}

The Optical Gravitational Lensing Experiment (Udalski et al. 1992) is
devoted to search for dark matter in our galaxy using microlensing
phenomena. As a by-product of this program an extensive photometry of
variable stars in the Galactic Bulge has been accumulated (Udalski et al.
1994, 1995, 1996, 1997). We have performed a systematic frequency analysis
of RR~Lyrae and $\delta$~Scuti stars of this sample. The results for
monoperiodic variables are presented elsewhere (Poretti 2001). Here, we
discuss the main properties of the identified multiperiodic pulsators.

\section{Search for multiperiodicity}

The OGLE-1 database contains 215 RR Lyrae and 53 $\delta$~Scuti variables.
The data span $\sim\! 900$\thinspace days, with typically 130-150 I-band
measurements per star.  As a first step of analysis, the lightcurve is
fitted with the Fourier sum of the form

$${\rm m_I}(t) = {\rm A}_0 + \sum_{i=1}^N {\rm A}_i \cos ({2\pi\over {\rm\, P}_{\! 1}} t + \phi_i)$$

\noindent with primary pulsation period ${\rm P}_{\! 1}$ being adjusted as well. 
Next, the search for secondary periodicities is performed. Two different 
metods are applied:

\smallskip

\noindent a) we compute Fourier power spectrum of the residuals of the fit 

\noindent b) we supplement the fitting formula with additional cosine term 
             with a trial period ${\rm P}_{\! 2}$. We then fit the data for 
	     different values of ${\rm P}_{\! 2}$, searching for the value 
	     that reduces the dispersion of the fit in a significant way.

\smallskip

\noindent Both methods yield the same results. As a third step, the Fourier 
fit with two identified frequencies and their detectable linear combinations 
is performed. The search for additional periodicities is then repeated. The
process is stopped when no significant term appears.

\section{Canonical double-mode pulsators}

We have identified 6 "normal" double-mode variables, five among
$\delta$~Scuti stars and one (BW7~V30) among RR~Lyrae stars. They are listed
in Table\thinspace 1.  The period ratios are very typical and correspond
to the ratios of the first two radial overtones. The only exception is
BWC~V82 with ${\rm P}_{\! 2}/{\rm P}_{\! 1} = 0.55$, where a higher order
radial mode (third overtone) has to be invoked. 

\begin{table}
\caption{Canonical double-mode variables}
\tabcolsep=15pt
\begin{center}
\begin{tabular}{lccc}
\hline\\[-10pt] Star & & ${\rm P}_{\! 1}$\,[day] & ${\rm P}_{\! 2}$/${\rm P}_{\! 1}$ \\[2pt] \hline\\[-10pt]
\object{BW2 V142} & & 0.066 & 0.778 \\
\object{BW9 V192} & & 0.076 & 0.754 \\
\object{BW1 V207} & & 0.085 & 0.774 \\
\object{BW1 V109} & & 0.106 & 0.774 \\
\object{BWC V82}  & & 0.161 & 0.550 \\[1pt]\hline\\[-10pt]
\object{BW7 V30}  & & 0.362 & 0.743 \\[1pt]\hline\\[-20pt]
\end{tabular}
\end{center}
\end{table}

\vskip -3truecm

\section{Blazhko effect: nonradial oscillations in RR~Lyrae stars}

In 37 RR Lyrae variables a different type of multiperiodicity is found --
additional peaks very close to the primary pulsation frequency are present.
Their amplitudes are usually below 0.06\thinspace mag.  The secondary
frequencies are well-resolved within our dataset and are not due to a
secular period variability. Their beating with the primary (radial)
pulsation results in an apparent long-term amplitude and phase modulation,
a phenomenon referred to as Blazhko effect. 

This new multiperiodic behaviour comes in two different flavours: we see
either a single secondary peak, forming a {\it doublet} with the primary
frequency (1~RRc and 27~RRab stars) or a pair of secondary peaks, which
together with the primary frequency form an {\it equidistant triplet}
centered on the primary peak (1~RRc and 8~RRab stars).

While the frequency triplet can result from periodic amplitude and/or phase
modulation of a purely radial pulsation, such a process cannot produce a
doublet. The observed period ratios (${\rm P}_{\! 2}/{\rm P}_{\! 1}=0.95 -
1.02$) are not compatible with excitation of two radial modes.  The
unavoidable conclusion is that a secondary component of the doublet
must correspond to a {\it nonradial mode of oscillation}.

In Table\thinspace 2 we present the statistics of variables with closely
spaced frequencies for the Galactic Bulge sample, compared with the numbers
derived for the LMC (Alcock et al.  2000; Welch et. al. 2002). In both
populations close doublets/triplets occur more frequently in
fundamental-mode pulsators than in overtone pulsators. Interestingly
enough, the fraction of multiperiodic RRab stars is two times higer in the
Galactic Bulge than in the LMC. It is tempting to speculate that the
difference in the incidence rate is related to difference in metallicity of
the two populations.  This hypothesis can be tested when photo\-metry of
RR~Lyrae stars in the SMC is analysed.

\begin{figure}
\begin{center}
\mbox{\epsfxsize=0.6\textwidth\epsfysize=0.41\textwidth\epsfbox{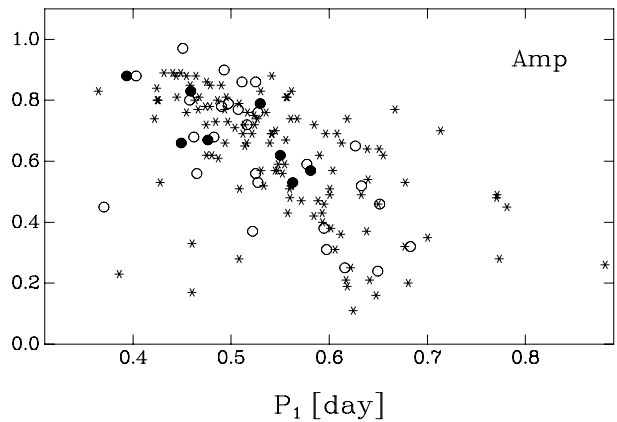}}
\caption{Amplitudes of Galactic Bulge RRab stars (as given by OGLE team). 
Asterisks\thinspace : monoperiodic variables, open circles\thinspace : 
va\-riables with frequency doublets, filled circles\thinspace : variables 
with frequency triplets.}
\end{center}
\end{figure}

\begin{table}
\caption{Incidence rate of RR Lyrae variables with closely spaced frequencies
in the Galactic Bulge and in the LMC. Standard deviations calculated 
assuming Poisson distribution.}
\tabcolsep=12pt
\begin{center}
\begin{tabular}{lcccc}
\\[-20pt]
\hline\\[-10pt] Type & Galactic Bulge & LMC \\[2pt] \hline\\[-10pt]
fundamental-mode pulsators (RRab) &   $23.2\pm 3.9$\% &  $ 10.2\pm 0.8$\% \\
overtone pulsators (RRc)          & ~$\,3.1\pm 2.2$\% & ~$\,3.9\pm 0.5$\% \\[1pt]\hline\\[0pt]
\end{tabular}
\end{center}
\end{table}

With the sample of 35 multiperiodic RRab stars, we can discuss the group
properties of this type of variables.  In Fig.\thinspace 1 we plot
amplitudes of the Galactic Bulge RRab stars as a function of their
periods.  The presence of secondary frequencies has no apparent effect on
the pulsation amplitude. We note that close frequency doublets are
detected with roughly the same probability at all periods represented in
this sample. The occurrence of triplets, on the other hand, seems to be
limited to ${\rm P} < 0.6$\thinspace day.

Fig.\thinspace 2 shows the frequency separation $\Delta {\rm f} = 
{\rm f}_2 - {\rm f}_1$ (${\rm f} = 1/{\rm P}$) for multiperiodic
RRab variables identified in this work. For 80\% of all doublets
$\Delta {\rm f}$ is positive, corresponding to secondary frequency being
{\it higher} than the primary one.  Identical distribution of 
$\Delta {\rm f}$ has also been found for RRab stars in the LMC (Welch
et al.  2002). The negative values of $\Delta {\rm f}$ occur only in a
narrow range of periods between 0.49\thinspace day and 0.53\thinspace day.
Curiously, this particular period range seems to be avoided by stars with
triplets. 

The frequency separation in the triplets tends to be slightly smaller than
in the doubles.  For both triplets and doublets the separation 
$\Delta {\rm f}$ is significantly smaller than in the overtone RR~Lyrae
variables of the LMC (Alcock et al.  2000).

\begin{figure}
\begin{center}
\mbox{\epsfxsize=0.63\textwidth\epsfysize=0.41\textwidth\epsfbox{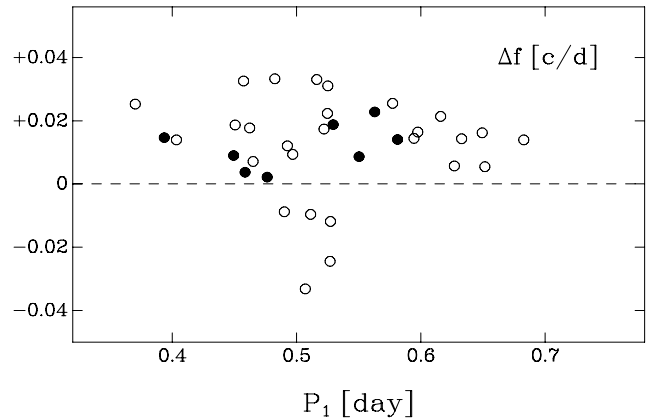}}
\caption{Frequency difference, $\Delta {\rm f}$, for Galactic Bulge 
multiperiodic RRab stars. Symbols the same as in Fig.\thinspace 1.}
\end{center}
\end{figure}

\bigskip

\noindent{\bf Acknowledgements.} This work has been supported by Polish 
KBN grants 2~P03D~002~20 and 5~P03D~012~20.

\vfill

\end{document}